# Collaborative thesaurus tagging the Wikipedia way *

Jakob Voss

Wikimedia Deutschland e.V.

jakob.voss@wm.sieheauch.de

## ABSTRACT
This paper explores the system of categories that is used to classify articles in Wikipedia. It is compared to collaborative tagging systems like del.icio.us and to hierarchical classification like the Dewey Decimal Classification (DDC). Specifics and commonalities of these systems of subject indexing are exposed. Analysis of structural and statistical properties (descriptors per record, records per descriptor, descriptor levels) shows that the category system of Wikimedia is a thesaurus that combines collaborative tagging and hierarchical subject indexing in a special way.

## Categories and Subject Descriptors
H.3.1 [**Information Systems**]: Information Storage and Retrieval – *Content Analysis and Indexing*

## 1. INTRODUCTION
For decades subject indexing has been a common practise within library and information science that is only practised by librarians and documentalists Now a similar process is gaining popularity on the web: collaborative tagging systems allow users to publicly annotate links, photographs, references and other items with keywords or 'tags'. Popular annotation systems like del.ico.us, flickr, technorati, RawSugar, CiteULike, etc. have already reached several million users. They are part of a movement with several web-based services like weblogs, wikis, and peer-to-peer that focus on the communication and interaction between users instead of using the traditional centralized one-to-many model of publication. One of the most popular social software sites is Wikipedia, an open content encyclopaedia that is collaboratively edited by its users. By the end of 2005 there are Wikipedias in more than 200 languages with 2.8 million articles combined. The goal of Wikimedia Foundation (Wikipedia's parent organization) is to provide free access to all human knowledge of the world; beside Wikipedia there is a free media archive (Wikicommons), a repository of free e-books (Wikibooks), a source of News reports (Wikinews), a multilingual dictionary and thesaurus (Wiktionary) and other projects. All of them are collaboratively developed by its users using the MediaWiki software. The content is free under the GNU Free Documentation License (GFDL), meaning that it may be freely used, edited, copied, and redistributed subject to the restrictions of that license. MediaWiki provides so called categories that are used to classify articles and other pages in the Wikimedia projects. Assigning categories to Wikipedia articles is a form of collaborative tagging with some particularities. Namely the category system is a thesaurus with hierarchical relationships between tags and categories can both added and removed. In this paper I will present Wikipedia's category system and compare it with other systems of collaborative tagging (del.icio.us), classification (Dewey Decimal Classification) and thesaurus indexing (Medical Subject Headings). The analyses are supported with statistics of general structural properties of classifications, collaborative tagging systems and thesauri.

## 2. WIKIPEDIA'S CATEGORY SYSTEM
From the beginning of Wikipedia there have been efforts to categorize its articles. The continuing growth of Wikipedia by number of articles was accompanied with a growth of lists and lists of lists in which participants tried to sort articles by topic. In May 2004 a new category system was introduced with version 1.3 of the MediaWiki software. At this time the English Wikipedia had around 280,000 articles and 3,000 regular contributors.[1] Soon they started to experiment with the new feature and the question of how to use categories in the right way was discussed intensively. Tagging was mostly unknown (Joshua Schachter started del.icio.us in late 2003, in May 2004 it had around 400,000 post while this number is added in one week by the end of 2005).[2] Users were used to hierarchical systems and I was very sceptical about categories and the way they were implemented in MediaWiki.[3] Page titles in MediaWiki are composed of two parts: an optional namespace name, and the remainder of the title. For example, the article about Apples has a discussion page with the title [[Discussion:Apple]] (double brackets mark page titles in MediaWiki) so it is in the 'Discussion' namespace while [[Apple]] without a namespace prefix is in the main namespace. Categories are pages in the 'Category' namespaces. A page can be put in a category by adding a category tag with a link to the category page. For example, the article [[Apple]] contains the tag [[Category:Agriculture]] and the tag [[Category:Apples]]. At the bottom of each page all assigned categories are listed with links to the category page where an automatic index of all pages tagged with this category is shown. This view is read-only: to add or remove a page from a specific category the article has to be edited. Possibilities to rename and reorganise categories are limited: when a category is renamed with a redirect from the old name to the new name, tags in the articles remain unchanged and pages still show up under the old category name. Nevertheless many members of the Wikipedia community spend their time in removing adding and category tags. At January 2, 2006 the English Wikipedia contained 923,196 articles (94% of them with at least one category) and 91,502 categories. Both grow exponen-tially (see figure 1) while categories increase faster (8.1% per month) than articles (6.5% per month). Since categories are also used for general maintenance (for instance to mark short articles and to index other namespace's pages, one cannot directly deduce the growth in categories per page.

---

[1] http://en.wikipedia.org/wikistats/EN/TablesWikipediaEN.htm

[2] See http://deli.ckoma.net/stats (January 1, 2005) and http://lists.del.icio.us/pipermail/discuss/2004-May/000353.html

[3] See my mail 'Categories considered harmful' from June 19, 2004 at the wikipedia-l@wikimedia.org mailing list: http://mail.wikipedia.org/pipermail/wikipedia-l/2004-June/033490.html

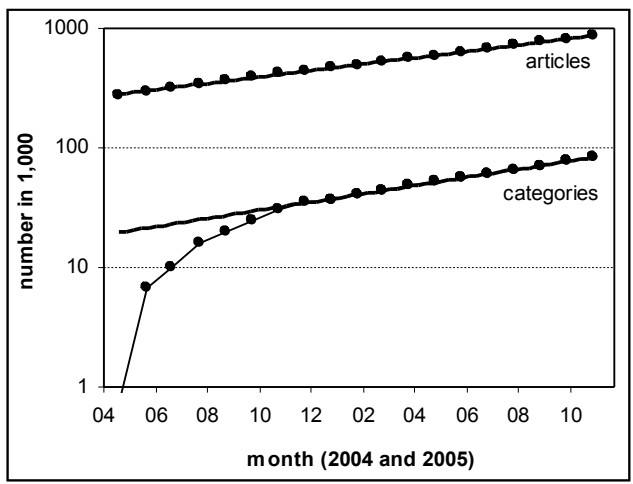

**Figure 1. Articles and categories in the English Wikipedia both growth exponentially**

An essential difference between known collaborative tagging systems and Wikipedia's categories is that one can also assign categories to other categories. This way hierarchical relationships with supercategories and subcategories are defined. From these hierarchies one can derive tree structures like those of known classifications. Most of the categories are connected to a a selected main category that is superordinated to all other categories. Table 1 shows the position of the descriptor for earth's moon in Dewey Decimal Classification (DDC) and English Wikipedia.

**Table 1. position of the descriptor for earth's moon for DDC and Wikipedia. DDC is fixed but there are multiple ways in which 'Moon' is classified in Wikipedia.**

| Dewey Decimal Classification (DDC) | | Wikipedia Categories |
|---|---|---|
| 5 | Science | Science |
| 59 | Earth sciences & geology | Astronomy |
| 559 | Other parts of world and extraterrestrial worlds | Astronomical objects |
| 559.9 | Extraterrestrial worlds | Moons |
| 559.91 | Earth's moon | Moon |

Erroneously you could think of Wikipedia's category system as a special kind of classification. But the category path from Science to Moon is only one possible hierarchy. Every category can be assigned to many other categories, while there is only one parent class for each class in a classification. The category system of Wikipedia is not a classification but a thesaurus – a classical tool of information retrieval. Wikipedia categories and their relations don't have strict semantics like known ontologies in computer science. However semantic information can be derived: [6] using co-occurrences of categories to calculate similarities between categories and create a "semantic map" of English Wikipedia.

## 3. THESAURI IN INFORMATION RETRIEVAL AND WIKIPEDIA

A thesaurus is a controlled vocabulary of terms that can be used as keywords. The terms are connected via relations to find the best fitting term. The word 'thesaurus' comes from Greek and first appeared in English in 1736 (see [1]). In a broader sense it then meant 'a treasury or storehouse of knowledge, as a dictionary, encyclopaedia and the like.' Peter Mark Roget's famous *Thesaurus of English Words and Phrases* (1852) introduced the concept of a linguistic thesaurus. In this special kind of dictionary words are arranged systematically to let authors find synonyms, phrases or more precise terms. Many people may know this type of thesaurus from their word processor that provides a simple thesaurus as writing aid. In the context of information retrieval a thesaurus has a slightly different focus: in an information retrieval thesaurus concepts are arranged instead of terms. Its purpose is to provide a controlled vocabulary where every tag ('descriptor') has a marked meaning. In information retrieval the term thesaurus was first used by Peter Luhn in 1957. It evolved in the 1950s when several kinds of subject indexing arose. Around 1950 the Uniterm system was developed by Mortimer Taube. Its basic idea was to use a limited set of equally important terms and index documents with combination of them. For instance a document about stomach cancer is indexed with 'stomach' and 'cancer'. This strategy is known as postcoordination while in precoordination a single subject term 'stomach cancer' is used. To solve classical problems of indexing, that are homonyms ('cancer' can also be a crab) and synonyms ('cancer', 'carcinoma', 'neoplams'), precise rules for vocabulary control and term relationships were invented. The current international standard for thesauri ISO 2788 defines a thesaurus as 'the vocabulary of a controlled indexing language, formally organized so that the a priori relationships between concepts are made explicit'. There are three kinds of relationships in thesauri:

a) the equivalence relationships
b) the hierarchical relationship
c) the associative relationship

Each relationship between term A and B has a corresponding relationship from term B to term A. The standard relationships and abbreviations as defined in ISO 2788 [7] and ANSI/NISO Z39.19 [2] are listed in table 2. The concept of narrower terms and broader terms has direct counterparts in the MediaWiki software. Equivalence relationships could be defined with redirects between categories but the current version (1.7alpha) does not distinguish between preferred terms and non-preferred terms (redirects) when categories are assigned to pages. Associations between categories are possible with normal links between category pages but in contrast to thesaurus relationships these links are not symmetric by default.

**Table 2. Standard relationships in a thesaurus**

| Relationship | Abbreviation | MediaWiki counterpart |
|---|---|---|
| Equivalence (synonymy) | USE<br>USE FOR | Redirects |
| Hierarchy | **B**roader **T**erm<br>**N**arrower **T**erm | Categories of a category<br>Subcategories of a category |
| Association | **R**elated **T**erm | (Links between category pages) |

An example of the relations between terms in a thesaurus is shown in figure 2. It shown the class for stomach cancer ('Stomach Neoplasms') in the 2005 version of the Medical Subject Headings (MeSH) and all of its broader terms. MeSH is a controlled vocabulary for indexing journal articles and books in the life sciences. The corresponding structure in Wikipedia is shown in figure 3. There is no category about stomach cancer but an article is assigned to two categories. The graph of all broader categories is not that detailed but more extensive because Wikipedia is not limited to life sciences. The category structure above 'Medicine' is less straightforward. This is because general concepts like 'Science' and 'Nature' are not that easy to structure and it shows that the Wikipedia category system evolves from bottom to top. Most thesauri are like MeSH and limited to a specific area but there are also some universal subject headings and thesauri, namely the Library of Congress Subject Headings (LCSH). One cannot rely on the name of an indexing system: subject headings can be thesauri or flat lists and one of the most know 'thesaurus', the Getty thesaurus of geographical names is in fact a classification. A simple measure is the number of relations per descriptor that is one for a classification, less for unconnected lists of terms and higher for thesaurus-like structures.

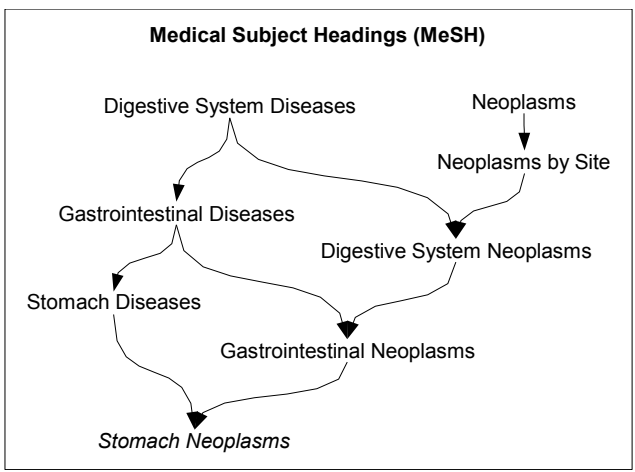

**Figure 2. Example from the MeSH thesaurus**

Another common mistake refers to the meaning of hierarchical relationships. Hierarchy seems to have a strict semantic that does not fit to the vagueness of the world. In practise there are always several ways to classify an object (for instance diseases by region of the body, dangerousness, type of treatment etc.). A faceted classification is not a principal solution of monohierarchy because independent 'facets' have to be separated. If one uses poly-hierarchy like in a thesaurus, the system is much more flexible but complexity increases and ambiguities have to be solved. In object-oriented programming the diamond problem can occur when multiple inheritance is allowed: given that every vehicle has a maximum velocity and there are water vehicles and land vehicles – do amphibious vehicles have one maximum velocity or two? To build a formal ontology that can fully processed automatically ambiguity must be avoided by all means. But human beings can judge in individual cases. This is why 'broader term' and 'narrower term' do not need to be defined more precisely in many cases (some thesauri differ types of hierarchical relationships but they normally don't have strict semantics like designed in the vision of 'Semantic Web').

## 4. COLLABORATIVE TAGGING SYSTEMS

The field of collaborative tagging on the web is growing enormously and most insights about it are directly published and discussed on the web (see [4] for one of the first traditional scientific papers about it). Thomas Vander Wal, who also coined the term 'folksonomy' as a combination of 'folk' and 'taxonomy', defined two types of folksonomies [11]: broad folksonomies and narrow folksonomies. Broad folksonomies arise if many people publically tag the same items mainly for themselves. The most popular example of a broad folksonomy is del.icio.us where tags are used to describe public bookmarks. A narrow folksonomy, on the other hand, is the result of a smaller number of individuals, tagging items mostly for later personal retrieval. An example of a narrow folksonomy is flickr where people tag their photos they have made. In both cases some popular tags are used very frequently and most tags are less known and used. The resulting distribution of tag frequency is a power law curve with a long tail.

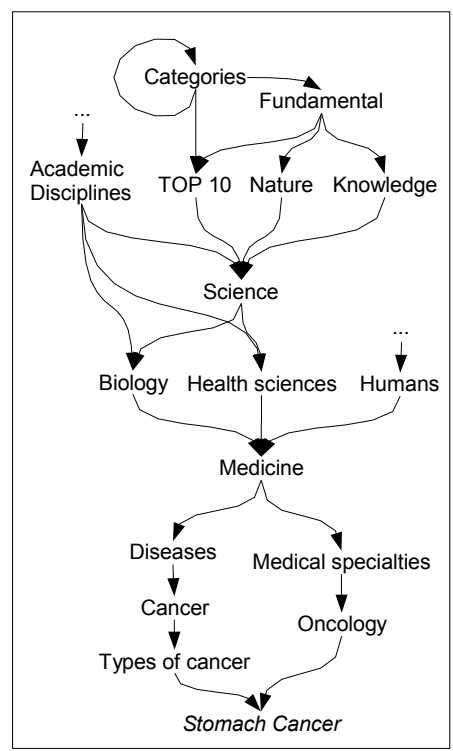

**Figure 3. The article about stomach cancer and its categories and broader categories in the English Wikipedia**

Power law distributions (also referred to as Pareto distribution and Zipfian distribution), have been found in many systems like word frequencies, file sizes and the in- and out-degree distribution for the Web. In Wikipedia Voss [14] detects power law distributions for hyperlink degrees, authors per article, articles per author, and edits per author. Zipf's version of the law states that the size $y$ of the $x$'th largest object in a set is inversely proportional to its rank. On a log-log scale the relation $y = x^{-\lambda}$ where $\lambda$ is constant is linear (see figure 5). Mitzenmacher's overview [9] shows that power law and lognormal distributions are very similar and from a pragmatic point of view one can interchange them in some cases.

## 5. STRUCTURAL PROPERTIES

To analyse commonalities and differences between Wikipedia categories, flat collaborative tagging systems and classifications, I measured and compared some of their structural properties. The data of the English Wikipedia was taken from a copy of the database at January 2, 2006 at 0:30 UTC. Dewey Decimal Classification statistics are taken from [12]. The data of del.icio.us was kindly provided by Phillipp Keller who set up a page with statistics that are generated from the RSS feed of the most recent del.icio.us posts (http://deli.ckoma.net/stats). Between August 1 and December 31, 2005 half a million posts were grabbed and analysed. The popularity of tags is based on a sample during week 2 and 3, 2006. The most popular tags in flickr and Millionsofgames, a directory of games, are accessible at the their websites. Both were grabbed at January 7, 2006. There is also a special MediaWiki page that lists the most popular categories (Special:Mostlinkedcategories). Additionally I tested some Wikipedias in other languages to confirm result.

### 5.1 Descriptors per record

While in classification every document is classified with only one (primary) class, a tagging allows assigning multiple descriptors. Compared to professional indexing there is no directive on how many descriptors to assign to a document in collaborative tagging. Figure 4 shows the distribution of the number of descriptors per record for del.icio.us (tags per post) and Wikipedia (categories per page). They both follow an exponential distribution. The chosen region with between 1 and 9 tags covers 94% of all Wikipedia articles (99% skipping the 6% of articles without any category) and 98% of all del.icio.us posts in the sample. For the small number of items with 10 tags or more the distribution merges into a power law (with exponent –6 for Wikipedia and –4 for del.icio.us). In general the percentage $p$ of records tagged with $n$ descriptors is $p(n) = \lambda e^{-\lambda n}$ with mean parameter $\lambda = 0.5$ for del.icio.us and $\lambda = 0.6$ for Wikipedia. In detail only the numbers for less then 3 tags differ: Wikipedia has less pages with 1 and more pages with 2 categories. Skipping pages without any category only the second deviation remains: it's slightly more common than expected to assign exactly two categories to an article, for unknown reasons.

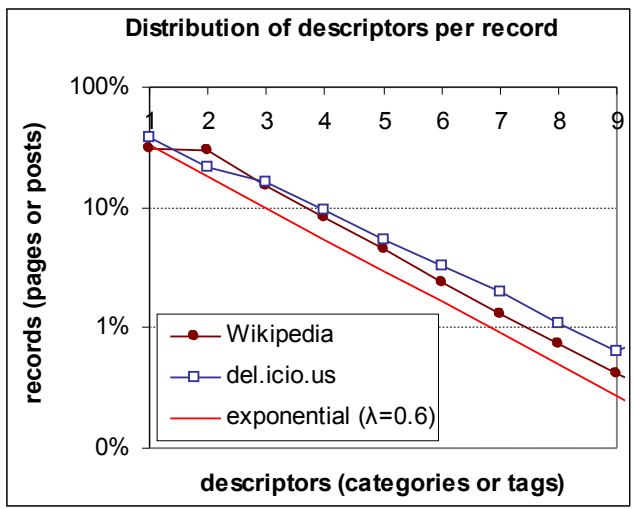

**Figure 4. The number of descriptors assigned to a record is distributed exponentially for low numbers (<10)**

The number of broader terms per category in Wikipedia is distributed in a same way but with $\lambda = 0.4$ (including 87% of all categories that have supercategories) and a long power law tail for categories with 10 or more supercategories (appendix 8.3).

### 5.2 Records per descriptor

It has been pointed out by different people [10,5] that in collaborative tagging the popularity of tags is distributed by a power law at least for selected records. This distribution also applies for collaborative tagging in general: Figure 5 shows the popularity of the 25 most used tags in several tagging systems. The distribution shows itself to be linear in log-log scale so one can derive a power-law distribution. The exponents determined by simple linear regression are $\lambda=0.35$ (Flickr), $\lambda=0.96$ (English Wikipedia), $\lambda=0.94$ (DDC), $\lambda=0.46$ (del.icio.us), $\lambda=0.59$ (millionsofgames). Tests with the cumulative distribution at the other side of the tail (number of categories with 1,2,… pages) show that at least Wikipedia category sizes follow a power law. However for other languages the value can vary between 0.75 and 1.5. For a more detailed analysis the full category data should be included. The exponent differs between Wikipedia and DDC on the one side and flat tagging systems on the other side. In hierarchical tagging descriptors seem to be distributed more steeply.

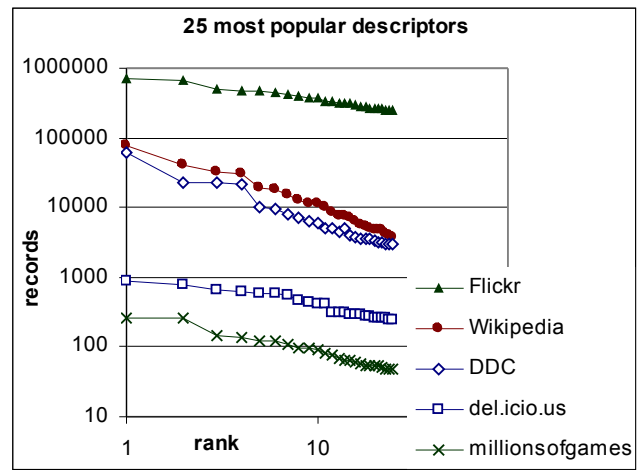

**Figure 5. The 25 most used tags in several tagging systems are distributed similarly**

### 5.3 Descriptor levels

Both classifications and thesauri provide hierarchical relations between descriptors. Normally there is one or a set of root descriptors without broader terms ('top terms' in a thesaurus) from which one can reach all other descriptors from. The level of a descriptor can be defined as the length of a minimal path of hierarchical relations between the descriptor and a top term. In a monohierarchical classification there is only one path for each descriptor because of its tree structure. Figure 5 shows the distribution of descriptor levels in DDC compared to the English Wikipedia. In Wikipedia the category 'categories' (see figure 2) was also declared as top term. The DDC numbers are taken from [12] including additional built class numbers from the WordCat database. A virtual top term that all 10 basic classes are assigned to was assumed. Both levels are distributed normally, confirmed by a KS-Test with p > 0.99. Also mean (DDC: 5.7; Wikipedia: 5.4) and standard derivation (DDC: 1.36; Wikipedia: 1.27) have similar values. Tests with Wikipedias in other languages show that some thousand categories are needed to get a normal distribution; mean varies between 4.0 and 5.5, and standard derivation varies between 0.9 and 1.7. More hierarchical systems of different kinds have to be analysed to determine if this distributions are typical for large hierarchies.

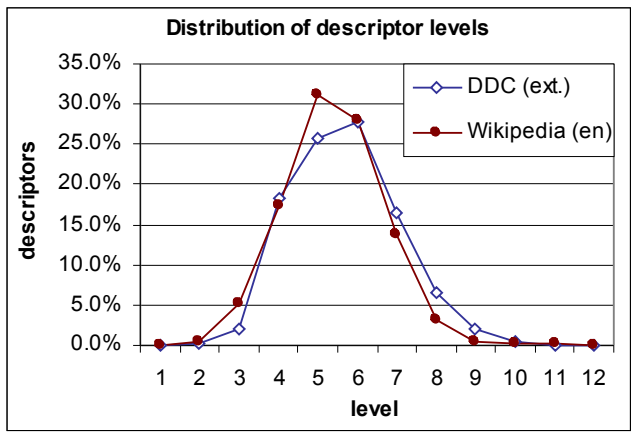

**Figure 6. Distribution of levels of DDC classes and Wikipedia categories (DDC: m=5.7, σ=1.36; Wikipedia: m=5.4, σ=1.27)**

## 6. SUMMARY AND CONCLUSIONS

In this paper three indexing systems were discussed: classifications (or taxonomies), simple index terms, and thesauri (see figure 7). Classifications are widely used to sort and structure. A popular example is the Dewey Decimal Classification (DDC). In many cases you can assign strict semantics to the simple hierarchy of a classification (ontologies). In context of collaborative tagging classification is often used to show its limits in contrast to 'tagging'[10]. Tagging with uncontrolled keywords is gaining popularity on the web in collaborative tagging (or folksonomy) systems. The tags in these systems are not connected (beside their correlation) and they are used without specific rules – nevertheless powerful structures evolve from collaborative action. The third system is a thesaurus. In a thesaurus tags are connected more flexible with less strict semantics.

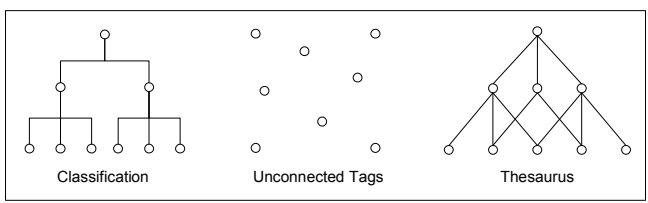

**Figure 7. Structure of indexing systems**

It is shown that Wikipedia's category system is a thesaurus that is collaboratively developed and used for indexing Wikipedia articles. Thesauri are known in information retrieval for around 50 years but far less known and used as hierarchical classification and flat indexing. Collaborative thesaurus creation and tagging is a new method of information retrieval that combines thesauri and collaborative tagging. Beside Wikipedia it has not been performed nor analysed before. The structural properties of classification (DDC), flat collaborative tagging (del.icio.us) and thesaurus indexing (Wikipedia) are compared in table 3. If multiple descriptors can be associated with a record, the number is distributed exponentially with a power law tail for the small number of heavily tagged records. Exponential distribution parameters have similar values of 0.6 in Wikipedia (pages per category) and 0.5 in del.icio.us (tags per post) for up to 9 descriptors. The number of supercategories per category in Wikipedia is also distributed exponentially with λ = 0.4. The popularity of descriptors (records per tag) is distributed by power laws for all analyzed indexing systems with a more steep spread in hierarchical systems. In a hierarchical system each descriptor has a level related to a top term. Levels are distributed normally both in Wikipedia and DDC with mean around 5.5. A broad implication is that all indexing systems share typical distributions of tags, records and level – instead of confrontating collaborative tagging and indexing by experts you should consider the conceptual properties of the different indexing systems presented in this paper. Also the research of library and information science and information retrieval should be regarded.

**Table 3. Distribution of structural parameters of classification (DDC), thesaurus (Wikipedia) and flat collaborative tagging (del.icio.us)**

|  | DDC | del.ico.us | Wikipedia |
|---|---|---|---|
| tags per record | 1* | distributed exponentially with a power law tail ||
| broader terms per term | 1 (tree) | 0 (no hierarchy) | distributed exponentially |
| levels | distributed normally | 1 (no hierarchy) | distributed normally |
| records per tag | power law distribution |||

\* but you can also use more than one class

Further analyzes should include full data sets and more indexing systems for instance the full MeSH thesaurus. Moreover dynamics of creation and modification of tags and their usage can be explored. To complete and refine the results, detailed comparisons of classifications, simple index terms, and thesauri on the one hand and controlled indexing and collaborative tagging on the other hand are needed. In practice mapping and conversion between tagging systems and their instances are needed, but a very complex issue [3]. You can also use Wikipedia articles to tag other resources [8], and directly connect them with semantic relations [13] instead of their categories. Guy and Tinkin [5] note the problem that folksonomies aim to purposes: the personal collection, and the collective collection. In Wikipedia this is not the case because a specific category is only assigned to a specific article once for all users. Maybe this is essential for collaborative thesaurus tagging, but other systems are also thinkable. For a broader application the implementation in MediaWiki lacks simple methods to rename and reorganize categories as well as symmetric see-also-relations and functionalities to import and export terminologies for instance in RDF. Nevertheless it contains simple synonymity control and hierarchical relations that other folksonomies do not provide. Further research should review traditional and new systems, find out their main characteristics and get the best system for each application.

## 7. ACKNOWLEDGEMENTS

For feedback and support I would like to thank Philipp Keller, Philipp Mayr, Joseph Reagle, and my submission's unknow reviewers for Collaborative Web Tagging Workshop, WWW2006

## 8. REFERENCES

[1] Aitchison, Jean; Dextre Clarke, Stella: *The Thesaurus - A Historical Viewpoint, with a Look to the Future*. In: Cataloging & Classification Quarterly 37, 3/4 (2004) 5-21.

[2] ANSI/NISO Z39.19-2003: *Guidelines for the Construction, Format and Management of Monolingual Thesauri*.

[3] Doerr, Martin: *Semantic Problems of Thesaurus Mapping*. In: Journal of Digital Information 1, 8 (2001). http://jodi.ecs.soton.ac.uk/Articles/v01/i08/Doerr/

[4] Golber, Scott A.; Huberman, Bernando A.: *The Structure of Collaborative Tagging Systems*. In: Journal of Information Science 32, 2, 198-208. http://arxiv.org/abs/cs/0508082

[5] Guy, Marieke; Tonkin, Emma: *Folksonomies – Tidying up Tags?*. In: D-Lib Magazine 12, 1 (2006). http://www.dlib.org/dlib/january06/guy/01guy.html

[6] Holloway, Todd; Bozicevic, Miran; Börner, Katy: *Analyzing and Visualizing the Semantic Coverage of Wikipedia and Its Authors*. (to appear in 2006). http://arxiv.org/cs.IR/0512085

[7] ISO 2788:1986 *Guidelines for the establishment and development of monolingual thesauri*

[8] Linksvayer, Mike: *Going overboard with Wikipedia tags* (12 January 2006) http://gondwanaland.com/mlog/2006/01/12/wikipedia-tags/

[9] Mitzenmacher, Michael: *A Brief History of Generative Models for Power Law and Lognormal Distributions*. In: Internet Mathematic 1, 2, 226-251 (2001).

[10] Shirky, Clay: *Ontology is Overrated: Categories, Links, and Tags* (May 2005). http://shirky.com/writings/ontology_overrated.html

[11] Vander Wal, Thomas: *Explaining and showing broad and narrow folksonomies* (21 February 2005). http://www.personalinfocloud.com/2005/02/

[12] Vizine-Goetz, Diane: *Classification Schemes for Internet Resources Revisited*. In: J. of Internet Cataloging 5, 4 (2002)

[13] Völkel, Max; Krötzsch, Markus; Vrandecic, Denny; Haller, Heiko: *Semantic Wikipedia*. In: Proceedings of the 15th international conference on World Wide Web (2006). http://www2006.org/programme/item.php?id=4039

[14] Voss, Jakob: *Measuring Wikipedia*. In: Proceedings of the 10th International Conference of the ISSI (2005). http://eprints.rclis.org/archive/00003610/

## 9. APPENDIXES

### 8.1 Number of tags per item

| Tags | Wikipedia | | Del.icio.us sample | |
|---|---|---|---|---|
| 0 | 52264 | 6% | – | – |
| 1 | 281572 | 30% | 213755 | 38% |
| 2 | 277956 | 30% | 121639 | 22% |
| 3 | 141270 | 15% | 91254 | 16% |
| 4 | 77047 | 8% | 54255 | 10% |
| 5 | 41778 | 5% | 30916 | 6% |
| 6 | 22251 | 2% | 18199 | 3% |
| 7 | 11864 | 1% | 11422 | 2% |
| 8 | 6783 | 1% | 6146 | 1% |
| 9 | 3885 | 0% | 3546 | 1% |
| Sum | 916670 of 923196 | 99% | 551132 of 564330 | 98% |

### 8.2 Descriptor levels

| Level | DDC | Wikipedia |
|---|---|---|
| 0 | 1 (virtually) | 899 (most of them un-categorized categories) |
| 1 | 10 | 67 |
| 2 | 99 | 448 |
| 3 | 879 | 4915 |
| 4 | 8121 | 16753 |
| 5 | 11372 | 29914 |
| 6 | 12360 | 26992 |
| 7 | 7354 | 13141 |
| 8 | 2917 | 3043 |
| 9 | 923 | 435 |
| 10 | 248 | 241 |
| 11 | 36 | 128 |
| 12 | 11 | 55 |
| 13 | 2 | 7 |

### 8.3 Broader terms per term in Wikipedia

| Broader terms | Terms | Broader terms | Terms |
|---|---|---|---|
| 1 | 15572 | 6 | 1491 |
| 2 | 9143 | 7 | 1100 |
| 3 | 4219 | 8 | 895 |
| 4 | 2831 | 9 | 718 |
| 5 | 1997 | | |

## 8.4 Most popular tags

| | flickr | | Wikipedia | | DDC | | del.icio.us | | millionsofgames | |
|---|---|---|---|---|---|---|---|---|---|---|
| | tag | count | categories | count | class | count | tag | count | tag | count |
| 1 | wedding | 700583 | GFDL images | 77769 | 813.54 | 60578 | blog | 902 | simple | 262 |
| 2 | party | 648072 | Disambiguation | 41825 | 811.54 | 23009 | software | 780 | fun | 260 |
| 3 | family | 506812 | Public domain images | 32166 | 929.20973 | 22252 | design | 661 | space | 148 |
| 4 | friends | 467551 | Redirects from US postal abbreviation | 29971 | 823.914 | 21958 | reference | 619 | 3d | 140 |
| 5 | japan | 466556 | Logos | 18693 | 813.52 | 9903 | web | 601 | funny | 122 |
| 6 | travel | 441749 | Album covers | 18073 | 823912 | 9360 | programming | 589 | cute | 119 |
| 7 | vacation | 413476 | User-created public domain images | 15128 | 843.914 | 8004 | music | 562 | animals | 110 |
| 8 | christmas | 397056 | Fair use images | 12903 | 863 | 7283 | tools | 472 | addictive | 99 |
| 9 | london | 375612 | Screenshots of movies and television | 11474 | 823 | 6261 | news | 447 | rpg | 95 |
| 10 | nyc | 364353 | 1911 Britannica | 11043 | 861 | 5966 | linux | 409 | multiplayer | 91 |
| 11 | beach | 329385 | Free use images | 10217 | 821.914 | 5102 | art | 404 | shockwave | 82 |
| 12 | birthday | 327222 | User en | 8695 | 822.33 | 4978 | CSS | 318 | sports | 78 |
| 13 | me | 314712 | NowCommons | 7727 | 823.8 | 4439 | Blogs | 317 | fighting | 70 |
| 14 | california | 313853 | Promotional images | 7707 | 812.54 | 5104 | photography | 314 | flash | 64 |
| 15 | trip | 312039 | Screenshots of computer and video games | 6982 | 510 | 4068 | howto | 299 | addicting | 64 |
| 16 | cameraphone | 294374 | Redirects from UN/LOCODE | 6457 | 641.5 | 3678 | webdesign | 296 | water | 60 |
| 17 | nature | 282541 | Uploader unsure of copyright status | 5652 | 841.914 | 3625 | video | 291 | football | 56 |
| 18 | art | 273209 | User en-N | 5200 | 821 | 3499 | mac | 280 | shooting | 55 |
| 19 | summer | 266324 | Articles to be merged | 5111 | 895.135 | 3486 | shopping | 276 | match three | 55 |
| 20 | paris | 265264 | American actors | 4877 | 759.13 | 3444 | web2.0 | 267 | kids | 53 |
| 21 | sanfrancisco | 264452 | United States government images | 4862 | 811.52 | 3141 | tutorial | 263 | grouper | 53 |
| 22 | france | 263531 | Album stubs | 4656 | 843.912 | 3134 | ajax | 260 | strategy | 52 |
| 23 | italy | 254069 | Film actors | 4136 | 248.4 | 3029 | java | 253 | tetris | 49 |
| 24 | newyork | 245418 | Articles lacking sources | 4077 | 081 | 2987 | buisness | 251 | brain teaser | 49 |
| 25 | china | 244897 | Book covers | 3720 | 833.914 | 2943 | opensource | 248 | reactions | 48 |

**Sources**

Flickr: http://www.flickr.com/photos/tags/ (January 7, 2006)
Wikipedia: http://en.wikipedia.org/wiki/Special:Mostlinkedcategories (January 7, 2006)
DDC: [12] (Appendix I, page 10)
del.icio.us: Phillipp Keller
Millionsofgames: http://www.millionsofgames.com/keywords/ (January 10, 2006)